\documentclass[aps,showpacs,prep,graphics]{revtex4}
\usepackage{amssymb}
\usepackage[dvips]{graphicx}
\usepackage{amsmath}
\usepackage{bm}
\usepackage{epsfig}

\begin{document}

\title{ Interacting quintessence from new formalism of gravitoelectromagnetism formulated on a geometrical scalar-tensor gauge theory of gravity.}

\author{ $^{1}$ Jos\'e Edgar Madriz Aguilar\thanks{E-mail address: madriz@mdp.edu.ar} and  $^{2}$ M. Montes
\thanks{E-mail address: mariana.montnav@gmail.com} }
\affiliation{$^{1}$ Departamento de Matem\'aticas, Centro Universitario de Ciencias Exactas e ingenier\'{i}as (CUCEI),
Universidad de Guadalajara (UdG), Av. Revoluci\'on 1500 S.R. 44430, Guadalajara, Jalisco, M\'exico.  \\
and\\
$^{2}$ Centro Universtario de los Valles\\
Carretera Guadalajara-Ameca Km. 45.5, C.P. 46600, Ameca, Jalisco, M\'exico.\\
E-mail:  jose.madriz@academicos.udg.mx, 
madriz@mdp.edu.ar,
 mariana.montes@academicos.udg.mx}

\begin{abstract}
 We derive an interacting quintessence model on the framework of a recently introduced new class of geometrical scalar-tensor theories of gravity formulated on a Weyl-Integrable geometry, where the gravitational sector is described by both a scalar and a tensor metric field. By using a Palatini variational principle we construct a scalar-tensor action invariant under the Weyl symmetry group of the background geometry, which in the Einstein-Riemann frame leads to a gravitoelectromagnetic theory. We use the gauge freedom of the theory and the fact that the Weyl scalar field couples with matter fields to formulate an interacting quintessential model with a non-canonical kinetic term, where the quintessence field has a geometrical origin. Due to this non-canonicity we obtain that the mass of the quintessence field in the past epochs results to be small enough not to cause modifications in the baryon to photon ratio  during nucleosynthesis. 
\end{abstract}

\maketitle

\vskip .5cm
 Weyl-Integrable geometry, gravitoelectromagnetism, scalar-tensor gravity, quintessence, accelerated cosmic expansion, dark energy.

\section{Introduction}

Since the supernovae type Ia observations in 1998, the idea that the universe is expanding in an accelerating manner has become one of the main open problems in modern cosmology \cite{rquin1, rquin2, rquin3, rquin4}. Dark energy and modified gravity have been in general the proposals to explain the acceleration in the present epoch. As it is well-known the simplest model that fits the observational data is the $\Lambda$CDM model. However, it suffers from the cosmological constant and the coincidence problems. Moreover,  there are some problems of the $\Lambda$ CDM model  at the level of phenomenology of galaxies, as for example the missing satellite problem,  the cusp-core problem, and the too-big-too-fail problem \cite{bquin1, bquin2, bquin3, bquin4, bquin5}.  Quintessence models have arised as an attempt to alleviate the problems of the $\Lambda$CDM model. However, the main problem with many of these models is that so fine-tunning initial conditions are required in order to the mass of the quintessence field not provoke a variation in the baryon to photon ratio during nucleosynthesis, for example.  The cosmic coincidence problem has been the motivation of a particular kind of quintessence models known as interacting quintessence models. In this approach the fact that the dark energy density today is comparable with the present matter energy density has motivated the idea that dark energy may be coupled with dark matter. In fact, one of the problems in this kind of models is to find a physical motivation for such an interacting term.  In the literature we can find several alternatives to address the present accelerated expansion issue. Among these proposals we can count k-essence models \cite{rquin5, rquin6, rquin7, rquin8}, dark fluid models \cite{df1, df2, df3, df4, df5}, modified theories of gravity and Brane-World models \cite{rquin9, rquin10, rquin11, rquin12}, among many others. However, in this letter one of our interests is to investigate the present accelerated expansion issue in the context of non-Riemannian geometries. In particular on the class known as Weyl Integrable Geometry.
\\

In view that general relativity does not incorporate the Mach's principle of inertia, the scalar-tensor theories of gravity where introduced as an attempt to include this principle in a gravitational theory \cite{c11, c12}. However, one of the first problems was about the nature of the scalar field. For some researches it is not so clear if this scalar field plays the role of gravity or matter \cite{cont}. Another feature of this kind of theories is that there are two mathematical frames to make physical descriptions: the Jordan and the Einstein frames. Of course, a question that naturally emerges is about which of the both frames is the physical one. In the literature we can find authors that believe that the Jordan frame is the physical one and some others point that the Einstein frame is the physical one \cite{cont}. However, it is important to note that this controversy appears in scalar-tensor theories of gravity formulated on Riemannian geometrical backgrounds.  If now the background geometry turns non-Riemannian, this controversy is alleviated. Recently, a new kind of scalar-tensor theories of gravity was introduced, where the background geometry is not fixed apriori. Instead it is adopted the Palatini variational principle to obtain a more natural and appropriated geometry for the kind of scalar-tensor theories of gravity \cite{c10,c13,c14}. In this new approach both the scalar field and the metric tensor describe gravity, disappearing in this manner the ambiguity about the nature of the scalar field. Moreover, the controversy of the frames does not appear in here due to the invariance group of symmetries of the background geometry, which in this case is the Weyl-Integrable. \\

The Weyl geometry is a particular class of non-riemannian geometries that for different reasons, and even when it suffers from the second clock effect pointed by Einstein, in the last years has increased the interest of theoretical physicists \cite{c1,c2,c3,c4,c5,c6}.  The Weyl geometry has been adopted as  geometrical framework to propose for example, extensions of the standard model of particles physics and general relativity \cite{c7}, to formulate scale invariant theories of gravity coupled to the standard model of particles \cite{c7p}, it has been also used to extend Gauss-Bonnet gravities \cite{c8}, and to formulate gauge invariant theories of gravity \cite{c9}, among many other applications. The Weyl-Integrable geometry is a modified Weyl geometry that has arised as an attempt to avoid the second clock effect of the original Weyl geometry. This theory incorporates naturally a scalar field in the affine structure of the space-time. In recent investigations it has been shown that this scalar field can be related with the scalar field employed in traditional scalar-tensor theories of gravity, when the Riemannian geometry is replaced by a Weyl-Integrable one \cite{c10}. As we have mentioned above, in the few past years a new class of scalar-tensor theories has arised in which the motivation for the introduction of the scalar field is purely geometric \cite{c10,c13,c14}. In these approaches, the scalar field arises from a Weyl-integrable space-time geometry. The relevance on this particular geometry relies in much in the fact that general relativity can also be formulated in the language of Weyl-Integrable geometry \cite{c15,c16}. Moreover, cosmological and inflationary scenarios have been found, which seems to indicate that the physical inflaton field in the early universe could  be modeled by the Weyl scalar field \cite{c14}. The Weyl-Integrable geometry has been also investigated in the light of (2+1) gravity models, offering new options to give dynamics to the gravitational field \cite{c17}. \\

In this letter our interest is to formulate an interacting quintessence model on the framework of this new class of scalar-tensor theories of gravity developed in a geometrical background described by a Weyl-Integrable geometry, in which the action shares the same symmetries of the Weyl group of transformations. In order to do so, we propose a new action for this kind of theories that is invariant under both the diffeomorfism and the Weyl groups of transformations at the same time. The letter is organized as follows. In section I, we give a little introduction. In section II, we develop the general formalism in Weyl-Integrable geometry, by means of the introduction of an invariant action under Weyl-transformations. In section III, we show how to obtain gravitoelectromagnetism as view by a class of observers defined in the so called Einstein-Riemann frame. In section IV, we obtain the field equations in the presence of a matter field action. In section V, we formulate, as an application of the formalism, an interacting quintessence model, studying different dominance regimens for the quintessential dark energy component. Finally in section VI, we give some final remarks.

\section {The formalism in Weyl-Integrable geometry}

 Let us start considering an action for a scalar-tensor theory of gravity in the Jordan frame, which in vacuum is given by
\begin{equation}\label{f1}
S=\frac{1}{16\pi}\int d^{4}x\sqrt{-g}\left\lbrace \Phi {\cal R}+\frac{\tilde{\omega}(\Phi)}{\Phi}g^{\mu\nu}\Phi_{,\mu}\Phi_{,\nu}-\tilde{V}(\Phi)\right\rbrace ,
\end{equation}
where ${\cal R}$ denotes the Ricci scalar, $\tilde{\omega}(\Phi)$ is a function of the scalar field $\Phi$ and $\tilde{V}(\Phi)$ is a scalar potential. The action (\ref{f1}) can be rewritten in terms of the redefined field $\varphi=-\ln (G\Phi)$ in the form \cite{c14}
\begin{equation}\label{f2}
S=\int d^{4}x\sqrt{-g}\left\lbrace e^{-\varphi}\left[\frac{{\cal R}}{16\pi G}+\frac{1}{2}\omega (\varphi)g^{\mu\nu}\varphi_{,\mu}\varphi_{,\nu}\right]-V(\varphi)\right\rbrace,
\end{equation}
where we have made the identifications $(1/2)\omega(\varphi)=(16\pi G)^{-1}\tilde{\omega}[\varphi(\Phi)]$ and $V(\varphi)=(16\pi)^{-1}\tilde{V}(\varphi(\Phi))$. Now, we adopt the Palatini variational procedure to obtain the field equations, and thus the variation with respect to the affine connection yields \cite{c10}
\begin{equation}\label{f3}
\nabla _{\mu}g_{\alpha\beta}=\varphi_{,\mu}g_{\alpha\beta}.
\end{equation}
This equation corresponds exactly with the well known non-metricity condition for a Weyl-Integrable geometry. It implies that $\nabla_{\mu}$ is the covariant derivative defined in terms of the Weyl-Integrable connection. Thus, in order to distinguish this covariant derivative from the Riemannian one, from now on we will denote it as $^{(w)}\nabla$.  Hence, we can interpret that according to the Palatini variational principle the background geometry corresponding to the action (\ref{f2}) is the Weyl-integrable geometry. \\

One of the consequences of having a Weyl-Integrable background geometry is that the group of invariance of a scalar-tensor theory defined on this kind of geometry increases. This follows by studying  the symmetries of the condition (\ref{f3}). With this in mind, it can be easily seen that (\ref{f3}) results to be invariant under the transformations
\begin{eqnarray}\label{f4}
\bar{g}_{\alpha\beta}&=&e^{f}g_{\alpha\beta}\\
\label{f5}
\bar{\varphi}&=&\varphi +f,
\end{eqnarray}
where $f=f(x^{\alpha})$ is a well defined function of the spacetime coordinates.
In order to avoid misunderstandings it is important to note that the invariance of (\ref{f3}) is achieved when the both transformations (\ref{f4}) and (\ref{f5}) are applied at the same time. Thus, we have not just a simple conformal transformation, as it is the case for example in some theoretical approaches where the Weyl connection is obtained from the Levi-Civita connection by using a conformal transformation of the metric. It is worth mentioning that in fact a pure conformal transformation does not preserve the background metric.   In view of the above considerations it is not difficult to see that for the action (\ref{f2}) to formally describe a scalar-tensor theory of gravity formulated on a Weyl-Integrable geometrical setting, it is necessary the requirement that such an action must be an inva\-riant not only under  the diffeomorfism group, but also under the Weyl transformations (\ref{f4})-(\ref{f5}). \\

However, a transformation of the kinetic term in (\ref{f2}) according to (\ref{f4}) and (\ref{f5}) gives
\begin{equation}\label{f6}
\sqrt{-\bar{g}}\,\bar{\omega}(\bar{\varphi})e^{-\bar{\varphi}}\bar{g}^{\mu\nu}\bar{\varphi}_{,\mu}\bar{\varphi}_{,\nu}=\sqrt{-g}\,\bar{\omega}(\bar{\varphi}) e^{-\varphi}g^{\mu\nu}\left(\varphi_{,\mu}\varphi_{,\nu}+\varphi_{,\mu}f_{,\nu}+f_{,\mu}\varphi_{,\nu}+f_{,\mu}f_{,\nu}\right),
\end{equation}
 showing in this manner that the action (\ref{f2}) is not invariant under  Weyl transformations. Thus, in order to construct an invariant action, we introduce the gauge covariant derivative
\begin{equation}\label{f7}
\varphi_{:\mu}=(\,^{(w)}\nabla_{\mu}+\gamma B_{\mu})\varphi,
\end{equation}
where $B_{\mu}$ is a gauge vector field and  $\gamma$ is a coupling constant introduced to have the correct physical units. \\ 
 Hence,  the simplest action for a scalar-tensor theory of gravity on Weyl-Integrable geome\-trical  background, results to be
\begin{equation}\label{f9}
{\cal S}=\int d^{4}x\sqrt{-g}\,e^{-\varphi}\left[\frac{{\cal R}}{16\pi G}+\frac{1}{2}\omega (\varphi) g^{\alpha\beta}\varphi_{:\alpha}\varphi_{:\beta}-e^{-\varphi}V(\varphi)\right],
\end{equation}
where the invariance under (\ref{f4}) and (\ref{f5}) of (\ref{f9}) requires that the vector field $B_{\mu}$, the function $\omega(\varphi)$ and the scalar potential $V(\varphi)$, must obey respectively the transformation rules
\begin{eqnarray}\label{f10a}
\bar{\varphi}\bar{B}_{\mu} &=& \varphi B_{\mu}-\gamma^{-1}f_{,\mu},\\
\bar{\omega}(\bar{\varphi})&=&\omega(\bar{\varphi}-f)=\omega(\varphi),\label{f10b}\\
\bar{V}(\varphi)&=& V(\bar{\varphi}-f)=V(\varphi).\label{f10c}
\end{eqnarray}
Writing the expression (\ref{f10a}) in terms of the redefined vector field $W_{\alpha}=\varphi B_{\alpha}$ we arrive to
\begin{equation}\label{zaz1}
\bar{W}_{\mu}=W_{\mu}-\gamma^{-1}f_{,\mu}.
\end{equation}
To include a dynamics for $W_{\mu}$ we extend the action (\ref{f9}) in the form
\begin{small}
\begin{equation}
{\cal S}=\int d^{4}x\sqrt{-g}\,e^{-\varphi}\left[\frac{{\cal R}}{16\pi G}+\frac{1}{2}\omega(\varphi)g^{\alpha\beta}\varphi_{:\alpha}\varphi_{:\beta}-V(\varphi)e^{-\varphi}-\frac{1}{4}H_{\alpha\beta}H^{\alpha\beta}e^{-\varphi}\right],\label{f14}
\end{equation}
\end{small}
where  $H_{\alpha\beta}=W_{\beta ,\alpha}-W_{\alpha ,\beta}$ is the field strength of the gauge boson field $W_{\mu}$. In terms of the gauge field  $B_{\alpha}$ we have $H_{\alpha\beta}=\varphi_{,\alpha}B_{\beta}-\varphi_{,\beta}B_{\alpha}+\varphi F_{\alpha\beta}$ being $F_{\alpha\beta}=B_{\beta,\alpha}-B_{\alpha,\beta}$. Something interesting is that if we rewrite the kinetic term in the action (\ref{f14}), terms of the form  $\omega(\varphi)W^{\mu}W_{\mu}$ will appear and hence  the field $W^{\mu}$ can be interpreted as a  massive bosonic vector field when $\omega(\varphi)=constant$. 

\section{Gravitoelectromagnetism in the Einstein-Riemann frame}

As it was established, the non-metricity condition (\ref{f3}) is invariant under the Weyl transformations (\ref{f4}) and (\ref{f5}). An interpretation of these transformations is that they lead from one Weyl frame $(M,g,\varphi, B_{\alpha})$ to another $(M,\bar{g},\bar{\varphi},\bar{B}_{\alpha})$. We refer with Weyl  frame to a differential manifold endowed with a metric tensor, an affine connection, a scalar field (part of the connection) and a gauge vector field such that the Weyl-Integrable spacetime geometry holds. However, it is very useful to note that for the particular choice $f=-\varphi$ we can define the effective metric $h_{\mu\nu}=\bar{g}_{\mu\nu}=e^{-\varphi}g_{\mu\nu}$ such that $\bar{\varphi}=0$. The interesting of this election is that in this case the condition (\ref{f3}) reduces to the effective Riemann metricity condition: $\nabla_{\lambda} h_{\alpha\beta}=0$, and this is why we will refer to this frame $(M,\bar{g},\bar{\varphi}=0,\bar{B}_{\alpha})=(M,h,\bar{B}_{\alpha})$, as to the Einstein-Riemann frame. We will use this terminology to differentiate it from the traditional Einstein frame employed in Jordan-Brans-Dicke scalar tensor theories, in view that in the former the geodesics are not preserved under conformal transformations, while in here the geodesics are Weyl invariant \cite{c14}.\\

Something remarkable is that in the Weyl frame the scalar field plays the role of a dilatonic geometrical scalar field. However, in the Einstein-Riemann frame the Weyl scalar field is not more part of the affine structure. Thus the Weyl scalar field is view from the Einstein-Riemann frame as a physical field. It means that when we pass from the Weyl to the Einstein-Riemann frame, the Weyl field pass from being a geometrical field to a physical field. In addition, once we are in the Einstein-Riemann frame the action needs to be invariant only under the diffeomorphism group, and it implies that the geometrical invariance requirement for the gauge vector field $B_{\mu}$ given by (\ref{zaz1}) is no more valid in this  frame.\\

 Thus, we can see that due to the change of geometry the scalar field $\varphi$ and the gauge vector field $B_{\mu}$ have different properties and  interpretations in different frames. So, it results convenient to distinguish the quantities defined on the Einstein-Riemann frame from the ones introduced in the Weyl frame. With this idea in mind, we will now on use $\phi$ and $A_{\mu}$ for the scalar field and the gauge vector field, respectively, in the Einstein-Riemann frame. \\

Once we have established some of the physical and geometrical differences between both frames, it is not difficult to show that the action (\ref{f14}) in the Einstein-Riemann frame acquires the form
\begin{equation}\label{Rie2}
{\cal S}=\int d^{4}x\sqrt{-h}\left[\frac{R}{16\pi G}+\frac{1}{2}\omega(\phi)h^{\alpha\beta}{\cal D}_{\alpha}\phi{\cal D}_{\beta}\phi-V(\phi)-\frac{1}{4}F_{\alpha\beta}F^{\alpha\beta}\right],
\end{equation}
where ${\cal D}_{\mu}=\nabla_{\mu}+\gamma A_{\mu}$,  the operator $\nabla_{\lambda}$ is denoting the Riemannian covariant derivative and $F_{\mu\nu}=A_{\nu ,\mu}-A_{\mu,\nu}$. Evidently, the action (\ref{Rie2}) is invariant under the diffeomorphism group as required by the background geometry in this frame. The contribution of $A_{\alpha}$ in the non-canonical kinetic term through the covariant derivative ${\cal D}_{\mu}$ and the last term in the action (\ref{Rie2}), suggest that $A_{\mu}$ can play the role of an electromagnetic potential. For this to be so, $A_{\mu}$ must transform according to  
\begin{equation}\label{yq1}
\overset{\smile}{A}_{\mu}=A_{\mu}-\gamma^{-1}\sigma_{,\mu},
\end{equation}
where $\sigma =\sigma (x^{\alpha})$. It is important to note that  contrary to what happens in usual models of gravitation in which electromagnetism is involved, we are not introducing an electromagnetic field in the action on a Riemannian geometry. Instead, it was the change of frame that leave us to the action (\ref{Rie2}), in which $A_{\mu}$ is just a vector field and the covariant derivative comes from imposing a Weyl symmetry in the Weyl frame and not for other reasons as for example a gauge invariance imposition. It is the form of the residual action (\ref{Rie2}) that allow us to interpret $A_{\mu}$ as the electromagnetic potential, however, this can still be interpreted simply as an arbitrary vector field.
To ensure the invariance of the action (\ref{Rie2}) under  (\ref{yq1}), the next internal symmetries must also hold
\begin{eqnarray}\label{yq2}
\overset{\smile}{\phi}&=&\phi e^{\sigma},\\
\label{yq2p}
\overset{\smile}{\omega}(\overset{\smile}{\phi})&\equiv& e^{-2\sigma}\omega(e^{-\sigma}\overset{\smile}{\phi})=\omega(\phi)\\
\label{yq2q}
\overset{\smile}{V}(\overset{\smile}{\phi})&\equiv& V(e^{-\sigma}\overset{\smile}{\phi})=V(\phi).
\end{eqnarray}
Once we have interpreted $A_{\mu}$ as the electromagnetic potential, 
 the action (\ref{Rie2}) can be extended by adding a source term for $A_{\mu}$ in the form 
\begin{equation}\label{yq3}
{\cal S}=\int d^{4}x\sqrt{-h}\left[\frac{R}{16\pi G}+\frac{1}{2}\omega(\phi)h^{\alpha\beta}{\cal D}_{\alpha}\phi{\cal D}_{\beta}\phi-V(\phi)-\frac{1}{4}F_{\alpha\beta}F^{\alpha\beta}-J^{\alpha}A_{\alpha}\right],
\end{equation}
where $J^{\mu}$ is a conserved current density. Thus, straithforward calculations show that the action (\ref{yq3}) leads to the field equations
\begin{eqnarray}\label{Rie6}
&& G_{\mu\nu}=-8\pi G \left[\omega(\phi){\cal D}_{\mu}\phi{\cal D}_{\nu}\phi-\frac{1}{2}h_{\mu\nu}\left(\omega(\phi)h^{\alpha\beta}{\cal D}_{\alpha}\phi{\cal D}_{\beta}\phi- 2V(\phi)\right)-\tau_{\mu\nu}^{(em)}\right]\\
&& \omega(\phi)\Box\phi+\frac{1}{2}\omega^{\prime}(\phi)h^{\mu\nu}{\cal D}_{\mu}\phi{\cal D}_{\nu}\phi-\gamma\omega^{\prime}(\phi)A^{\mu}\phi{\cal D}_{\mu}\phi+\gamma\omega(\phi)\nabla_{\mu}A^{\mu}-\gamma^2\omega(\phi)A^{\mu}A_{\mu}\phi+V^{\prime}(\phi)=0,\label{Rie7}\\
&& \nabla_{\mu}F^{\mu\nu}= J^{\nu}-\gamma\omega(\phi)h^{\mu\nu}\phi{\cal D}_{\mu}\phi,\label{Rie8}
\end{eqnarray}
with  $\Box =h^{\mu\nu}\nabla_{\mu}\nabla_{\nu}$ being the usual D'Alambertian operator, $\tau_{\mu\nu}^{(em)}=T_{\mu\nu}^{(em)}-h_{\mu\nu}J^{\alpha}\! A_{\alpha}$, with $T_{\mu\nu}^{(em)}=F_{\nu\beta}F_{\mu}\,^{\beta}-\frac{1}{4}h_{\mu\nu}F_{\alpha\beta}F^{\alpha\beta}$ being the energy-momentum tensor for a free electromagnetic field. In this manner, the field equations (\ref{Rie6}), (\ref{Rie7}) and (\ref{Rie8}) can be interpreted as they describe a theory of gravitoelectromagnetism on a Riemannian geometrical background. \\

In general, it is not an easy task to solve the system (\ref{Rie6}) to (\ref{Rie8}). In order to decouple the scalar field $\phi$ from the electromagnetic potential $A_{\mu}$, we can implement an election of gauge. In the case of  cosmological applications it is important to remember that the cosmological principle is not compatible with the existence of a vector field on large scales.  However, in some scenarios of the  very early universe, it may be possible from our approach to construct a model for cosmic structure formation in which the seeds of cosmic structure can evolve together with an electromagnetic part. In this models for example, the electromagnetic field can be considered on quantum small scales, as an electromagnetic perturbation. For the moment we leave these kind of applications for further investigation. We are now more interested in applications of the model to the present cosmic accelerating expansion epoch. Thus, as a cosmological application of the formalism here developed, in the next sections we will present an interacting quintessence model in order to invesigate if the Weyl scalar field as viewed in the Einstein-Riemann frame can play the role of a quintessence dark energy field. However, in order to motivate the introduction of a dark matter sector in the next section we will show how to incorporate matter in the present geometrical formalism.

\section{The field equations in presence of matter}

Until now, we have considered a geometrical scalar-tensor gauge theory of gravity in vacuum. However, to formulate an interacting quintessence dark energy model, first it turns out to be more convenient a little discussion about how an action for matter fields $S_m$ enters in the formalism.  As it was shown in \cite{matweyl}, due to the fact that gravity is in here described by the two geometric fields $g_{\alpha\beta}$ and $\varphi$, it is reasonable to expect both couple with matter in an arbitrary Weyl frame. A Weyl invariant action for matter sources can be written in the form \cite{matweyl}
\begin{equation}\label{wm1}
S_{m}=\int d^{4}x\sqrt{-g}\,e^{-2\varphi}L_{m}\left(e^{-\varphi}g_{\mu\nu},\Psi,^{(w)}\!\nabla\Psi\right),
\end{equation}
with $\Psi$ denoting some matter field and  $L_m$ representing the matter lagrangian which is constructed taking into account the prescription $L_{m}(g,\varphi,\Psi,^{(w)}\!\nabla\Psi)\equiv L_{m}^{(sr)}(e^{-\varphi}g,\Psi,^{(w)}\!\nabla\Psi)$, where $L_{m}^{(sr)}$ is denoting the lagrangian for the field $\Psi$ in the flat Minkowski space-time of special relativity. Hence , the energy-momentum tensor $T_{\mu\nu}(\varphi,g,\Psi, ^{(w)}\!\nabla\Psi)$ for matter sources in an arbitrary Weyl frame $(M,g,\varphi)$ is determined by
\begin{equation}\label{wm2}
\delta\int d^{4}x \sqrt{-g} e^{-2\varphi}L_{m}(\varphi,g_{\mu\nu}, \Psi,^{(w)}\!\nabla\Psi) = \int d^{4}x \sqrt{-g}e^{-2\varphi}T_{\mu\nu}(\varphi,g_{\mu\nu},\Psi,^{(w)}\!\nabla\Psi)\delta(e^{\varphi} g^{\mu\nu}),
\end{equation}
where $\delta$ denotes the variation with respect to both $g_{\mu\nu}$ and $\varphi$. In this manner, it is not difficult to see that the field equations in the Einstein-Riemann frame in presence of matter sources read
\begin{eqnarray}\label{wm3}
&& G_{\mu\nu}=-8\pi G T_{\mu\nu}-8\pi G \left[\omega(\phi){\cal D}_{\mu}\phi{\cal D}_{\nu}\phi-\frac{1}{2}h_{\mu\nu}\left(\omega(\phi)h^{\alpha\beta}{\cal D}_{\alpha}\phi{\cal D}_{\beta}\phi- 2V(\phi)\right)-\tau_{\mu\nu}^{(em)}\right]\\
&& \omega(\phi)\Box\phi+\frac{1}{2}\omega^{\prime}(\phi)h^{\mu\nu}{\cal D}_{\mu}\phi{\cal D}_{\nu}\phi-\gamma\omega^{\prime}(\phi)A^{\mu}\phi{\cal D}_{\mu}\phi+\gamma\omega(\phi)\nabla_{\mu}A^{\mu}-\gamma^2\omega(\phi)A^{\mu}A_{\mu}\phi+V^{\prime}(\phi)=0,\label{wm4}\\
&& \nabla_{\mu}F^{\mu\nu}= J^{\nu}-\gamma\omega(\phi)h^{\mu\nu}\phi{\cal D}_{\mu}\phi.\label{wm5}
\end{eqnarray}
It is important to note that because $T_{\mu\nu}(\varphi,g_{\mu\nu},\Psi,^{(w)}\!\nabla\Psi)$ and  $h_{\alpha\beta}=e^{-\varphi}g_{\alpha\beta}$, we can say that the Weyl scalar field couples with matter fields. For our purposes this is an important feature that allow us to propose an interacting quintessence model, in which the Weyl scalar field will play the role of a quintessence dark energy field and dark matter will be modeled by $T_{\mu\nu}$. In this manner, the geometrical coupling of the Weyl scalar field with matter motivates the idea that the both dark sectors may interact each other, which is one of the basic ideas of these kind of quintessence models.

\section{An interacting quintessence model}

In this section, we will derive as an application of the formalism a late-time cosmological scenario where the dark matter is described by the energy momentum tensor for matter $T_{\mu\nu}$ and the dark energy component is modeled by the geometrical Weyl scalar field $\phi$ in the Einstein-Riemann frame. The field $\phi$ will play the role of a quintessence field and an interaction with the dark matter sector will be allowed.  \\

In our model we adopt the cosmological principle and in order to implement it, we choose the gauge $\overset{\smile}{A}_{\mu}=0$, which corresponds to the election $\sigma_{,\mu}=\gamma A_{\mu}$, necessarily implies that in this gauge the electromagnetic part obeys $\overset{\smile}{F}_{\mu\nu}=0$. In this gauge the transformed action $\overset{\smile}{S}$ reads
\begin{equation}\label{quin1}
\overset{\smile}{S}=\int d^{4}x \,\sqrt{-h}\left[\frac{R}{16\pi G}+\frac{1}{2}\overset{\smile}{\omega}(\overset{\smile}{\phi})h^{\alpha\beta}\overset{\smile}{\cal D}_{\alpha}\overset{\smile}{\phi}\overset{\smile}{\cal D}_{\beta}\overset{\smile}{\phi}-\overset{\smile}{V}(\overset{\smile}{\phi})\right]+S_m,
\end{equation}
which reduces to
\begin{equation}\label{quin2}
\overset{\smile}{S}=\int d^{4}x\,\sqrt{-h}\left[ \frac{R}{16\pi G}+\frac{1}{2}\hat{\omega}(Q)h^{\mu\nu}Q_{,\mu}Q_{,\nu}-\hat{V}(Q)\right] +S_m,
\end{equation}
where we have denoted by $Q=\overset{\smile}{\phi}$ the quintessencial scalar field, $\hat{\omega}(Q)=\overset{\smile}{\omega}(\overset{\smile}{\phi})$ and $\hat{V}(Q)=\overset{\smile}{V}(\overset{\smile}{\phi})$. The field equations for (\ref{quin2}) are given by
\begin{eqnarray}\label{quin3}
&& G_{\mu\nu}=-8\pi G T_{\mu\nu}-8\pi G \left[\hat{\omega}(Q)Q_{,\mu}Q_{,\nu}-\frac{1}{2}h_{\mu\nu}\left(\hat{\omega}(Q)h^{\alpha\beta}Q_{,\alpha}Q_{,\beta}-\hat{V}(Q)\right)\right],\\
\label{quin4}
&&\hat{\omega}(Q)\Box Q+\frac{1}{2}\hat{\omega}^{\prime}(Q)h^{\mu\nu}Q_{,\mu}Q_{,\nu}+\hat{V}^{\prime}(Q)=0.
\end{eqnarray}
By defining the non-canonical effective energy-momentum tensor
\begin{equation}\label{eemt}
 T_{\mu\nu}^{(eff)}=T_{\mu\nu}+\left[\hat{\omega}(Q)Q_{,\mu}Q_{,\nu}-\frac{1}{2}h_{\mu\nu}\left(\hat{\omega}(Q)h^{\alpha\beta}Q_{,\alpha}Q_{,\beta}-\hat{V}(Q)\right)\right],
 \end{equation} 
 the conservation equation $\nabla_{\mu}T^{\mu\nu}_{(eff)}=0$ holds. On the other hand, according to the observational data the universe seems to be  3D spatially flat, thus we consider the FLRW line element
\begin{equation}\label{quin5}
ds^2=dt^2-a^2(t)(dx^2+dy^2+dz^2),
\end{equation}
where $a(t)$ is the cosmic scale factor. Now, in order to construct the model let us to implement the field transformation 
\begin{equation}\label{quin6}
\zeta=\int\sqrt{\hat{\omega}(Q)}dQ.
\end{equation}
The expression (\ref{quinuin6}) tranforms the non-canonical kinetic term in the action (\ref{quin2}) in a canonical term leading to the action
\begin{equation}\label{quin7}
\overset{\smile}{S}(\zeta)=\int\,d^{4}x\,\sqrt{-h}\left[\frac{R}{16\pi G}+\frac{1}{2}h^{\mu\nu}\zeta_{\mu}\zeta_{,\nu}-V_{eff}(\zeta)\right]+S_m ,
\end{equation}
where $V_{eff}(\zeta)=\hat{V}(\zeta(Q))$ is the redefined  effective potential. \\

Now, it follows from (\ref{quin3}), (\ref{quin4}), (\ref{quin5}) and (\ref{quin6}) that the independent field equations read
\begin{eqnarray}\label{quin8}
&& 3H^2=8\pi G (\rho_{\zeta}+\rho_m),\\
\label{quin9c}
&& \dot{\rho}_{\zeta}+\dot{\rho}_m+3H(\rho_{\zeta}+\rho_m+p_{\zeta})=0,
\end{eqnarray}
where $\rho_m$ is the energy density for dark matter, whereas the pressure and energy density for dark energy  are given by
\begin{eqnarray}\label{quin9}
p_{\zeta}=\frac{1}{2}\dot{\zeta}^2-V_{eff}(\zeta),\\
\label{quin9a}
\rho_{\zeta}=\frac{1}{2}\dot{\zeta}^2 + V_{eff}(\zeta).
\end{eqnarray}
If we assume that dark energy does not evolve independently of dark matter, the equation (\ref{quin9c}) yields
\begin{eqnarray}\label{quin10}
&&\dot{\rho}_m + 3H\rho_m=\eta(t),\\
\label{quin11}
&& \dot{\rho}_{\zeta}+3H(\rho_{\zeta}+p_{\zeta})=-\eta(t),
\end{eqnarray}
where $\eta(t)$ is an interaction function. This interaction between the both dark sectors motivates the relation $\rho_{m}(t)=\lambda(t)\rho_{\zeta}(t)$, with $\lambda(t)$ being a differentiable function that accounts for the time variation of the dark matter to dark energy ratio. Considering an equation of state parameter for the quintessence dark energy $\omega_{\zeta}=p_{\zeta}/\rho_{\zeta}$, with the help of equations (\ref{quin10}) and (\ref{quin11}) we obtain
\begin{equation}\label{quin12}
\dot{\rho}_{\zeta}+3H\left(1+\frac{\omega_{\zeta}}{1+\lambda}+\frac{\dot{\lambda}}{3H(1+\lambda)}\right)\rho_{\zeta}=0.
\end{equation}
Solving (\ref{quin12}) we arrive to 
\begin{equation}\label{quin13}
\rho_{\zeta}=\rho_{i}e^{-3\int_{t_{i}}^{t}\alpha(t')H(t')dt},
\end{equation}
where $\rho_{i}$ is some initial energy density and $\alpha(t)$ is an auxiliary function defined by 
\begin{equation}\label{quin14}
\alpha(t)=1+\frac{\omega_{\zeta}}{1+\lambda}+\frac{\dot{\lambda}}{3H(1+\lambda)}.
\end{equation}
Inserting (\ref{quin14}) in (\ref{quin11}) we obtain for the interaction function 
\begin{equation}\label{quin15}
\eta(t)=\left(\frac{\dot{\lambda}}{1+\lambda}-\frac{3H\lambda\omega_{\zeta}}{1+\lambda}\right)\rho_{\zeta}.
\end{equation}
On the other hand, in terms of $\lambda(t)$ the equation (\ref{quin8}) can be rewritten in the form
\begin{equation}\label{quin16}
H^2=\frac{8\pi G}{3}(1+\lambda)\rho_{\zeta}.
\end{equation}
In this manner, $\lambda>1$ corresponds to a matter dominance  epoch $\rho_m >\rho_{\zeta}$, when $\lambda=1$ we have an equality epoch $\rho_m=\rho_{\zeta}$ and $\lambda<1$ corresponds to a dark energy dominance epoch. According to this argument it is not difficult to see that $\lambda(t)$ needs to be a decreasing function of time.

\subsection{The dark energy dominance}

In this regimen it is satified the condition $\rho_{\zeta}\gg \rho_m$ or equivalently $\lambda(t)\ll 1$. If this is the case the equation (\ref{quin16}) reduces to
\begin{equation}\label{quin17}
H^2\simeq \frac{8\pi G}{3}\rho_{\zeta},
\end{equation}
and the energy density of dark energy (\ref{quin13}) reads
\begin{equation}\label{quin18}
\rho_{\zeta}\simeq \rho_{de}\left(\frac{a}{a_{de}}\right)^{-3(1+\omega_{\zeta})}e^{-(\lambda-\lambda_{de})},
\end{equation}
where $a_{de}=a(t_{de})$ and $\lambda_{de}=\lambda(t_{de})$, with $t_{de}$ being the time when the dark energy dominance regimen starts. Substituting (\ref{quin18}) in the Friedmann equation (\ref{quin17}) and imposing $a(t_{de})=a_{de}$ we obtain for solution
\begin{equation}\label{quin19}
a(t)\simeq a_{de}\left[1+\frac{3}{2}(1+\omega_{\zeta})\sqrt{\frac{8\pi G}{3}\rho_{de}}\int_{t_{de}}^{t}e^{-\frac{1}{2}(\lambda-\lambda_{de})}dt\right]^{\frac{2}{3(1+\omega_{\zeta})}}.
\end{equation}
Thus, with the use of (\ref{quin19}) the formula (\ref{quin18}) can be written as
\begin{equation}\label{quin20}
\rho_{\zeta}\simeq\rho_{de}\left[1+\frac{3}{2}(1+\omega_{\zeta})\sqrt{\frac{8\pi G}{3}\rho_{de}}\int_{t_{de}}^{t}e^{-\frac{1}{2}(\lambda-\lambda_{de})}dt\right]^{-2}e^{-(\lambda-\lambda_{de})}.
\end{equation}
Employing (\ref{quin9}) and (\ref{quin9a}) we arrive to
\begin{eqnarray}\label{quin21}
&& V_{eff}(t)\simeq \frac{1}{2}(1-\omega_{\zeta})\rho_{de}\left[1+\frac{3}{2}(1+\omega_{\zeta})\sqrt{\frac{8\pi G}{3}\rho_{de}}\int_{t_{de}}^{t}e^{-\frac{1}{2}(\lambda-\lambda_{de})}dt\right]^{-2}e^{-(\lambda-\lambda_{de})}.\\
\label{quin22}
&& \dot{V}_{eff}(t)\simeq -3(1+\omega_{\zeta})V_{eff}\left[1+\frac{3}{2}(1+\omega_{\zeta})\sqrt{\frac{8\pi G}{3}\rho_{de}}\int_{t_{de}}^{t}e^{-\frac{1}{2}(\lambda-\lambda_{de})}dt\right]^{-1}\sqrt{\frac{8\pi G}{3}\rho_{de}}e^{-\frac{3}{2}(\lambda-\lambda_{de})}-\dot{\lambda}V_{eff},\\
\label{quin23}
&& \zeta(t)\simeq \zeta_{de}+\sqrt{(1+\omega_{\zeta})\rho_{de}}\int_{t_{de}}^{t}\left[1+\frac{3}{2}(1+\omega_{\zeta})\sqrt{\frac{8\pi G}{3}\rho_{de}}\int_{t_{de}}^{t}e^{-\frac{1}{2}(\lambda-\lambda_{de})}dt\right]^{-1}e^{-\frac{1}{2}(\lambda-\lambda_{de})}dt.
\end{eqnarray}
In the limit case when $\lambda \to 0$ the equations (\ref{quin21}) to (\ref{quin23}) become
\begin{eqnarray}\label{quin24}
 && V_{eff}(t)\simeq \frac{1}{2}(1-\omega_{\zeta})\rho_{de}\left[1+\frac{3}{2}(1+\omega_{\zeta})\sqrt{\frac{8\pi G}{3}\rho_{de}}\,e^{\frac{1}{2}\lambda_{de}}(t-t_{de})\right]^{-2}e^{\lambda_{de}}.\\
 \label{quin25}
 && \dot{V}_{eff}(t)\simeq -3(1+\omega_{\zeta})V_{eff}\left[1+\frac{3}{2}(1+\omega_{\zeta})\sqrt{\frac{8\pi G}{3}\rho_{de}}\,e^{\frac{1}{2}\lambda_{de}}(t-t_{de})\right]^{-1}\sqrt{\frac{8\pi G}{3}\rho_{de}}\,\,e^{\frac{3}{2}\lambda_{de}},\\
 \label{quin26}
 && \zeta(t)\simeq \zeta_{de}+\sqrt{(1+\omega_{\zeta})\rho_{de}}\int_{t_{de}}^{t}\left[1+\frac{3}{2}(1+\omega_{\zeta})\sqrt{\frac{8\pi G}{3}\rho_{de}}\,e^{\frac{1}{2}\lambda_{de}}(t-t_{de})\right]^{-1}e^{\frac{1}{2}\lambda_{de}}dt.
\end{eqnarray}
An algebraic manipulation of (\ref{quin24}), (\ref{quin25}) and (\ref{quin26}), and the fact that $\dot{V}_{eff}=V_{eff}^{\prime}\dot{\zeta}$ lead to the quintessential potential
\begin{equation}\label{quin27}
V_{eff}(\zeta)\simeq V_{de}e^{\sqrt{24\pi G (1+\omega_{\zeta})}e^{\lambda_{de}}(\zeta_{de}-\zeta)}.
\end{equation}
Thus it can be obtained an exponential potential for $\zeta >\zeta_{de}$, in this limit case of the dark energy dominance epoch. In terms of the original quintesential field $Q$ the equation (\ref{quin27}) reads
\begin{equation}\label{quin28}
V_{eff}\simeq V_{de}\,e^{\sqrt{24\pi G (1+\omega_{\zeta})}e^{\lambda_{de}}(Q_{de}-\int \sqrt{\hat{\omega}(Q)}dQ)}.
\end{equation}
Therefore, we found that the exponential form for the effective potential $V_{eff}$ can be maintained just for some functional forms of $\hat{\omega}(Q)$. However, it not in general the case.

\subsection{The equality regimen}

In this regimen the condition to satisfy is now $\rho_{\zeta}=\rho_m$ or equivalently $\lambda(t)=1$. It means that there is an equality between the energy densities of the both dark components. In this case the equation (\ref{quin16}) yields
\begin{equation}\label{quin29}
H^2=\frac{16\pi G}{3}\rho_\zeta.
\end{equation}
The energy density for the quintessential field  (\ref{quin13}) under this approximation becomes
\begin{equation}\label{quin30}
\rho_{\zeta}\simeq \rho_{eq}\left(\frac{a}{a_{eq}}\right)^{-3(1+\frac{1}{2}\omega_{\zeta})}e^{-\frac{1}{2}(1-\lambda_{eq})},
\end{equation}
where $a_{eq}=a(t_{eq})$ and $\lambda_{eq}=\lambda(t_{eq})$, being $t_{eq}$ the time when the equality regimen begins. Substituting (\ref{quin30}) in the Friedmann equation (\ref{quin29}) and imposing $a(t_{eq})=a_{eq}$ the  solution for the scale factor can be written in the form
\begin{equation}\label{quin31}
a(t)\simeq a_{eq}\left[1+\frac{3}{2}(1+\frac{1}{2}\omega_{\zeta})\sqrt{\frac{16\pi G}{3}\rho_{eq}}\,e^{-\frac{1}{4}(1-\lambda_{eq})}(t-t_{eq})\right]^{\frac{2}{3\left(1+\frac{1}{2}\omega_{\zeta}\right)}}.
\end{equation}
Thus, with the use of (\ref{quin31}) the formula (\ref{quin30}) can be written as
\begin{equation}\label{quin32}
\rho_{\zeta}\simeq\rho_{eq}\left[1+\frac{3}{2}(1+\frac{1}{2}\omega_{\zeta})\sqrt{\frac{16\pi G}{3}\rho_{eq}}\,e^{-\frac{1}{4}(1-\lambda_{eq})}(t-t_{eq})\right]^{-2}.
\end{equation}
With the use of (\ref{quin9}) and (\ref{quin9a}) we arrive to
\begin{eqnarray}\label{quin33}
&& V_{eff}(t)\simeq \frac{1}{2}(1-\omega_{\zeta})\rho_{eq}e^{-\frac{1}{2}(1-\lambda_{eq})}\left[1+\frac{3}{2}\left(1+\frac{1}{2}\omega_{\zeta}\right)\sqrt{\frac{16\pi G}{3}\rho_{eq}}\,e^{-\frac{1}{4}(1-\lambda_{eq})}(t-t_{eq})\right]^{-2}e^{-(\lambda-\lambda_{eq})}.\\
\label{quin34}
&& \dot{V}_{eff}(t)\simeq -3\left(1+\frac{1}{2}\omega_{\zeta}\right)\sqrt{\frac{16\pi G}{3}\rho_{eq}}e^{-\frac{1}{4}(1-\lambda_{eq})}V_{eff}(t)\left[1+\frac{3}{2}\left(1+\frac{1}{2}\omega_{\zeta}\right)\sqrt{\frac{16\pi G}{3}\rho_{eq}}e^{-\frac{1}{4}(1-\lambda_{eq})}(t-t_{eq})\right]^{-1}\\
\label{quin35}
&& \zeta(t)\simeq \frac{2}{3}\left(\frac{1+\omega_\zeta}{1+\frac{1}{2}\omega_\zeta}\right)\rho_{eq}\left(\frac{16\pi G\rho_{eq}}{3}\right)^{\frac{1}{2}}e^{-\frac{1}{4}(1-\lambda_{eq})}\left[1+\frac{3}{2}\left(1+\frac{1}{2}\omega_{\zeta}\right)\sqrt{\frac{16\pi G}{3}\rho_{eq}}e^{-\frac{1}{4}(1-\lambda_{eq})}(t-t_{eq})\right]^{-1}\label{quin36}.
\end{eqnarray}
It follows from (\ref{quin34}), (\ref{quin35}) and (\ref{quin36}) that the quintessential potential written in terms of $\zeta$ is given by
\begin{equation}\label{quin39}
V_{eff}(\zeta)\simeq V_{eq}\,e^{-3\left(\frac{1+\frac{1}{2}\omega_\zeta}{1+\omega_\zeta}\right)\sqrt{\frac{16\pi G}{3}}(\zeta-\zeta_{eq})}.
\end{equation}
Thus for $\zeta >\zeta_{eq}$ it can be obtained an exponential potential.  In terms of the original quintesential field $Q$ the equation (\ref{quin39}) acquires the form
\begin{equation}\label{quin40}
V_{eff}\simeq V_{eq}\,e^{-3\left(\frac{1+\frac{1}{2}\omega_\zeta}{1+\omega_\zeta}\right)\sqrt{\frac{16\pi G}{3}}(\int\hat{\omega}(Q)dQ-Q_{eq})}.
\end{equation}
It follows from (\ref{quin40}) that similarly to the previous regimen the effective potential is depending of the $\hat{\omega}(Q)$ function.

\subsection{The matter dominance regimen}

In the case when $\rho_m\gg \rho_{\zeta}$ the matter component is dominant over the dark energy. Under this approximation, it is valid that $\lambda\gg 1$ and according to the equations  (\ref{quin13}) and (\ref{quin16}) the scale factor is given by
\begin{equation}\label{md1}
a(t)\simeq \left[a_{md}^{3/2}+\frac{3}{2}\sqrt{\frac{8\pi G \rho_{md}}{3}}\,(a_{md}\lambda_{md})^{1/2}(t-t_{md})\right]^{2/3},
\end{equation}
where $t_{md}$ is the time when the matter dominance epoch begins, $a_{md}=a(t_{md})$, $\rho_{md}=\rho(t_{md})$ and  $\lambda_{md}=\lambda(t_{md})$. Thus from (\ref{quin9}), (\ref{quin9a}) and (\ref{quin13}) we obtain
\begin{equation}\label{md2}
\rho_{\zeta}\simeq \rho_{md}a_{md}^3\left(\frac{\lambda_{md}}{\lambda}\right)\left[a_{md}^{3/2}+\frac{3}{2}\sqrt{\frac{8\pi G \rho_{md}}{3}}\,(a_{md}\lambda_{md})^{1/2}(t-t_{md})\right]^{-2}.
\end{equation}
Hence, the scalar field $\zeta$ reads
\begin{equation}\label{md3}
\zeta\simeq \zeta_{md}+\sqrt{1+\omega_{\zeta}}\,(\rho_{md}a_{md})^{1/2}\int_{t_{md}}^t\,\left(\frac{\lambda_{md}}{\lambda}\right)^{1/2}\left[a_{md}^{3/2}+\frac{3}{2}\sqrt{\frac{8\pi G \rho_{md}}{3}}\,(a_{md}\lambda_{md})^{1/2}(t-t_{md})\right]^{-1}dt.
\end{equation}
Notice that for $\lambda\gg 1$ the scalar field $\zeta$ tends to the constant $\zeta\simeq \zeta_{md}$, where $\zeta_{md}=\zeta(t=t_{md})$. For the potential in this regime we obtain
\begin{equation}\label{md4}
V_{eff}(t)\simeq \frac{1}{2}(1-\omega_{\zeta})\rho_{md}a_{md}^3\left(\frac{\lambda_{md}}{\lambda}\right)\left[a_{md}^{3/2}+\frac{3}{2}\sqrt{\frac{8\pi G \rho_{md}}{3}}\,(a_{md}\lambda_{md})^{1/2}(t-t_{md})\right]^{-2}.
\end{equation}
Then the potential in terms of $\zeta$ results to be
\begin{equation}\label{md5}
V_{eff}(\zeta)\simeq V_{md}\,\exp\left[-3\sqrt{\frac{8\pi G\rho_{md}}{3}}\,(a_{md}\lambda)^{1/2}(\zeta-\zeta_{md})\right].
\end{equation}
Therefore, the potential in terms of the quintessence field $Q$ becomes
\begin{equation}\label{md6}
V_{eff}(Q)\simeq V_{md}\,\exp\left[-3\sqrt{\frac{8\pi G\rho_{md}}{3}}\,(a_{md}\lambda)^{1/2}\left(\int\sqrt{\hat{\omega}(Q)} dQ-Q_{md}\right)\right].
\end{equation}
By using (\ref{md6}) the mass associated to $Q$ is given by
\begin{equation}\label{md7}
m_{eff}^2(Q)\simeq 24\pi G\, V_{md}(\rho_{md}a_{md})\lambda\,\hat{\omega}(Q_{md}).
\end{equation}
As it was expected, the mass for the quintessence field is also function of $\hat{\omega}(Q)$. It means that there exist a class of $\hat{\omega}(Q)$ functions that allow a small value of $m_{eff}$ when $t\to t_{md}$. This is important in order to not generate conflicts with the predictions of light element abundances during nucleosynthesis. In the next section we will give a more insight about this point.

\subsection{The attractor case}

As it is well known, in interacting quintessence dark energy models there exists a stationary attractor type solution characterized by a constant ratio between the both energy densities compatible with the present accelerating expansion of the universe \cite{AR1,AR2}. In our case this solution is achieved for constant $\lambda=\lambda_0$. In fact according to PlanckTT $+$ lowP $+$ lensing $+$ ext, can be  inferred that  $\lambda_0=0.44092\,^{+0.00508}_{-0.005695}$ \cite{obsp1,obsp2}. In this case the equation (\ref{quin14}) becomes
\begin{equation}\label{as1}
\rho_{\zeta}=\rho_{0}\left(\frac{a}{a_{0}}\right)^{-3\alpha_0},
\end{equation}
where $\alpha_0=1+\omega_{\zeta}/(1+\lambda_0)$, $\rho_0=\rho(t=t_0)$, $a_0=a(t=t_0)$ being $t_0$ the initial time when the accelerating expansion epoch begins. Inserting (\ref{as1}) in (\ref{quin16}) we obtain for the scale factor
\begin{equation}\label{as2}
a(t)=a_{0}\left[\frac{3\alpha_0}{2}\sqrt{\frac{8\pi G}{3}(1+\lambda_0)\rho_{0}}\,(t-t_{0})+1\right]^{\frac{2}{3\alpha_0}}.
\end{equation}
Thus the equation (\ref{as1}) yields
\begin{equation}\label{as3}
\rho_{\zeta}=\rho_{0}\left[1+\frac{3\alpha_0}{2}\sqrt{\frac{8\pi G}{3}(1+\lambda_0)\rho_{0}}\,(t-t_{0})\right]^{-2}.
\end{equation}
In this manner, with the help of (\ref{quin9}) and (\ref{quin9a}), the scalar field $\zeta$ and its potential read
\begin{eqnarray}\label{as4}
&& \zeta(t)=\frac{2}{3\alpha_0}\sqrt{\frac{3}{8\pi G}\left(\frac{1+\omega_{\zeta}}{1+\lambda_0}\right)}\,\ln\left[1+\frac{3\alpha_0}{2}\sqrt{\frac{8\pi G}{3}(1+\lambda_0)\rho_{0}}\,(t-t_{0})\right].\\
\label{as5}
&& V_{eff}(t)=\frac{1}{2}(1-\omega_{\zeta})\rho_{0}\left[1+\frac{3\alpha_0}{2}\sqrt{\frac{8\pi G}{3}(1+\lambda_0)\rho_{0}}\,(t-t_{0})\right].
\end{eqnarray}
Employing (\ref{as4}) and (\ref{as5}), the effective potential as a function of $\zeta$ is given by
\begin{equation}\label{as6}
V_{eff}(\zeta)=V_{0}\exp\left[-\alpha_0\sqrt{24\pi G\left(\frac{1+\lambda_0}{1+\omega_{\zeta}}\right)}\,(\zeta- \zeta_{0})\right].
\end{equation}
This potential as a function of the original quintessential field $Q$ reads
\begin{equation}\label{as7}
V_{eff}(Q)=V_{0}\exp\left[-\alpha_0\sqrt{24\pi G\left(\frac{1+\lambda_0}{1+\omega_{Q}}\right)}\,\left(\int\sqrt{\hat{\omega}(Q)}dQ-Q_{0}\right)\right].
\end{equation}
 It is important to note that as it happens in the previous cases, depending of the form of $\hat{\omega}(Q)$ the potential (\ref{as7}) not neccessarily obeys an exponential behavior. For example, if $\int\sqrt{\hat{\omega}(Q)}dQ=\ln(\xi Q)^{1/\xi}$, with $\xi$ being a dimensional constant, which corresponds to $\hat{\omega}(Q)=1/(\xi Q)$, the effective potential obeys $V_{eff}(Q)\simeq (\xi Q)^{-\beta/\xi}$, where the constant parameter $\beta=\alpha_0\sqrt{24\pi G\left[(1+\lambda_0)/(1+\omega_{\zeta})\right]}$. This is in fact a difference with respect to the Zimdahl and Pavon model \cite{AR1}, because the exponential behavior of the quintessential potential can change. The effective mass associated to the quintessential field $Q$ derived from the potential (\ref{as7}) results to be in our case
 \begin{equation}\label{as8}
 m_{eff}^{2}=V_{0}\alpha_0^2\,24\pi G\left(\frac{1+\lambda_0}{1+\omega_Q}\right)\hat{\omega}(Q).
 \end{equation}
 A common problem in models of quintesence is that the mass of the quintesence field results to be very large in the early epochs of the universe \cite{AR1,AR2}. Thus in view of the coupling of this field with matter, this large value of mass implies that the baryon to photon ratio determined by a successful nucleosynthesis period might change \cite{AR2}. In our model due to the non-canonical kinetic energy of the quintessential field $Q$, we have an extra degree of freedom that can alleviate this problem. If $t_{now}$ denotes the present time and $t_{past}$ denotes some time in the early universe, by means of the equation (\ref{as8}) we obtain the relation
 \begin{equation}\label{as8a}
 \frac{m_{eff}(t_{now})}{m_{eff}(t_{past})}=\frac{V(t_{past})}{V(t_{now})}\left( \frac{\alpha_{now}}{\alpha_{past}}\right)\sqrt{\frac{1+\lambda_{now}}{1+\lambda_{past}}}\,\frac{\hat{\omega}(Q_{now})}{\hat{\omega}(Q_{past})},
 \end{equation}
 where $\alpha_{past}$, $\lambda_{past}$ and $Q_{past}$ correspond to the values of $\alpha_0$, $\lambda_0$ and $Q$ in $t=t_{past}$, while $\alpha_{now}$, $\lambda_{now}$ and $Q_{now}$ denote the values of $\alpha$, $\lambda$ and $Q$ in $t=t_{now}$. Thus, as it is well known, to have a successful quintessence model we must require that $m_{eff}(t_{now})\simeq H_{now}$ and $m_{eff}(t_{past})=m_{*}$, with $m_{*}$ being small enough to avoid problems with nucleosynthesis. Thus, the expression (\ref{as8a}) yields
 \begin{equation}\label{as8b}
 \frac{H_{now}}{m_{*}}=\frac{V(t_{past})}{V(t_{now})}\left( \frac{\alpha_{now}}{\alpha_{past}}\right)\sqrt{\frac{1+\lambda_{now}}{1+\lambda_{past}}}\,\frac{\hat{\omega}(Q_{now})}{\hat{\omega}(Q_{past})}.
 \end{equation}
 Finally, the equation (\ref{as8}) can be written in terms of $m_{*}$ as
 \begin{equation}\label{as8c}
 m_{eff}^2=m_{*}^2\left(\frac{V_{now}}{V_{past}}\right)\left(\frac{\alpha_{now}}{\alpha_{past}}\right)\left(\frac{1+\lambda_{now}}{1+\lambda_{past}}\right)\frac{\hat{\omega}(Q_{now})}{\hat{\omega}(Q_{past})}.
 \end{equation}
 It follows from (\ref{as8c}) that for $t=t_{past}$, in particular in the nucleosynthesis epoch, $m_{eff}=m_{*}$, which is small enough to conciliate with the nucleosynthesis requiriments. Moreover, clearly using (\ref{as8b}), the equation (\ref{as8c}) for $t=t_{now}$ reduces to $m_{eff}(t_{now})\simeq H_{now}$. Something remarkable is that the smallness of $m_{*}$ can be achieved by considering apropiated $\hat{\omega}(Q)$ functions. Finally, the deceleration parameter in this case is given by
 \begin{equation}\label{aaau}
 q=\frac{1}{2}+\frac{3\omega_{Q}}{(1+\lambda_0)}.
 \end{equation} 
 According to the observational data: Planck + WP + BAO + SN, the present equation of state parameter  ranges in the interval $\omega_0=-1.10^{+0.08}_{-0.07}$ \cite{obsp1,obsp2}. Thus for $\lambda_0=0.44092$ the deceleration parameter (\ref{aaau}) take the values $-0.711<q<-0.5557$.

\section{Final Remarks}

In this letter we have formulated an interacting quintessence model in the framework, of a recently introduced \cite{c10,c13,c14}, geometrical scalar-tensor theory of gravity. We see that if we adopt a Palatini variational principle, it is obtained that the resultant background geometry for a scalar-tensor theory of gravity is the known as Weyl-Integrable and not the Riemannian one, as it is usual assumed in this kind of theories \cite{c10}.  In this geometry the non-metricity condition is invariant under the Weyl group of transformations (\ref{f4})-(\ref{f5}). However, we have shown that the action (\ref{f2}) for a scalar-tensor theory of gravity is not invariant under the Weyl group. Thus we have proposed the new simple invariant action for a scalar –tensor theory of gravity (\ref{f14}), in which, as it happens in quantum gauge theories, we have modified the covariant derivative by the introduction of the gauge vector field $B_{\mu}$.  As an extension of the results shown in \cite{matweyl}, we obtain that the transformations (\ref{f4}), (\ref{f5}) and (\ref{zaz1}) can be interpreted as they pass in general from one Weyl frame $(M,g,\varphi,B_{\mu})$ to another $(M, \bar{g},\bar{\varphi},\bar{B}_{\mu})$ in the same equivalence class. However under the election $f=-\varphi$ it is possible to obtain an effective Riemannian geometry by means of the introduction of the effective metric tensor $h_{\alpha\beta}=e^{-\varphi}g_{\alpha\beta}$. We call Einstein-Riemann frame to the set $(M,h, A_{\alpha})$ where $A_{\alpha}$ is the vector field $B_{\alpha}$ when we have implemented the metric redefinition $h$. It is important to make clear that what we have called Einstein-Riemann frame is different from the definition of Einstein frame in the usual scalar-tensor theories of gravity. In these theories the Einstein frame is defined as the frame in which the scalar field couples minimally to gravity \cite{Far}. Unlike the geometrical scalar-tensor theories here concerned, in the usual scalar-tensor theories the Einstein frame is introduced by a merely algebraic procedure because the background geometry in both frames (the Jordan and the Einstein frames) is the Riemannian geometry (see section III for a better explanation).  An interesting feature is that the scalar field in the Weyl frame is in fact part of the affine structure of the manifold and plays the role of a geometrical dilatonic field. However, the same field view in the Einstein-Riemann frame is a physical scalar field in virtue that it does not more form part of the affine structure. Due to the different groups of symmetry between the Weyl-Integrable and the Riemannian geometries, the Weyl scalar field must obey different symmetries in the two different frames, and it implies that the vector field also satisfy different gauge field transformations in both frames.  The symmetry transformation of the vector field in the Einstein-Riemann frame given by (\ref{yq1}) opens the possibility to interpret such vector field as the electromagnetic potential. Thus, we can interpret the field equations obtained in the Einstein-Riemann frame as they describe a theory of gravitoelectromagnetism in which both the scalar field and the electromagnetic potential have a geometrical origin.\\

Matter in this geometrical formalism is introduced by means of a Weyl invariant action which exhibits a coupling between the Weyl scalar field and the matter fields. This motivates the idea that if the scalar field plays the role of a quintessential scalar dark energy field then the dark energy sector may has an interaction with the dark matter sector. Thus, as an application of the formalism we have investigated the possibility that the Weyl scalar field can be viable to construct an interacting quintessence model to address the present cosmic accelerating expansion problem. Unlike the Zimdhal \& Pavon quintessence model, in here we have a non-canonical kinetic term determined by the $\hat{\omega}(\phi)$ function.  We have analized three different regimes of dominance: the dark energy dominance, the equality regimen and the matter dominance regimen, and the attractor case which corresponds to a constant dark matter to dark energy density ratio. We found that the effective potential $V_{eff}(Q)$ is given in terms of the $\hat{\omega}(Q)$ function.  Thus if $\hat{\omega}(Q)$ is constant we recover the typical exponential form of the potential obtained in some interacting quintessence models \cite{AR1,AR2}. However, if  for example 
$\hat{\omega}(Q)=1/(\xi Q)$, the effective potential has the form $V_{eff}(Q)\simeq (\xi Q)^{-\beta/\xi}$ with $\beta =\alpha_0\sqrt{24\pi G[(1+\lambda_0)/(1+\omega_{Q})]}$.  A typical problem in quintessence models is that the mass of the quintessence field results to be very large in the early epochs of the universe \cite{AR1,AR2}.  Thus, due to the coupling of both dark sectors such value of mass may cause for example variations in the baryon to photon ratio, changing the abundances of light elements during primordial nucleosynthesis \cite{AR2}.  This problem can be avoided in our model due to the function $\hat{\omega}(Q)$. In fact, the mass of the quintessence field $Q$ during nucleosynthesis results to be small enough to conciliate with the nucleosynthesis requirements by considering appropiated $\hat{\omega}(Q)$ function. Morever, the mass for $Q$ at the present time is $m_{eff}(t_{now})\simeq H_{now}$. Taking into account that observationally the present equation of state parameter  is given by  $\omega_0=-1.10^{+0.08}_{-0.07}$ \cite{obsp1,obsp2}, we obtain that the deceleration parameter for the attractor case ranges in the interval $-0.711<q<-0.5557$. In general, the theory of gravitoelectromagnetism obtained in the Einstein-Riemann frame can be the framework to study cosmic structure formation models in which the seeds of cosmic magnetic fields and of the cosmic structures can evolve in an unified manner. However we let this topic for further investigation.

\section*{Acknowledgements}

\noindent  J.E.Madriz-Aguilar and M. Montes  acknowledge CONACYT
M\'exico, Centro Universitario de Ciencias Exactas e Ingenierias and Centro Universitario de los Valles of Universidad de Guadalajara for financial support.

\end{document}